# Hydrodynamic pumping of a quantum Fermi liquid in a semiconductor heterostructure

J. J. Heremans[1], A. O. Govorov[1,2], D. Kantha[1] and Z. Nikodijevic[1]

[1]Ohio University, Nanoscale and Quantum Phenomena Institute, and Department of Physics and Astronomy, Athens, OH 45701, USA

[2]Institute of Semiconductor Physics, 630090 Novosibirsk, Russia

**Abstract.** We describe both experimentally and theoretically a hydrodynamic pumping mechanism in a Fermi liquid, arising from electron-electron interaction. An electron beam sweeping past an aperture is observed to pump carriers from this aperture. Experimentally, the pumping effect induces a current in the lead connected to the aperture, or induces a voltage signal corresponding to carrier extraction from the lead. Different geometries display the effect, and this work discusses one experimental geometry in detail. Theoretically, the solution of the Boltzmann equation, including an electron-electron collision integral, shows that the potential induced by injected electrons becomes *positive* in the regions nearby the main stream of injected electrons. Thus, the *repulsive* Coulomb interaction leads to an *attractive*, pumping force in the Fermi liquid. The pumping mechanism here described is shown to be qualitatively different from the Bernoulli pumping effect in classical liquids.

## 1. Introduction

Hydrodynamic effects can appear in charge transport through nanostructures, depending on the relative values of the momentum relaxation mean free path, the electron-electron interaction mean free path and the device dimensions [1-4]. In a high mobility two-dimensional electron system (2DES) at low temperatures, the electron-electron (*e-e*) interaction path, $l_{ee}$, can be longer than the momentum relaxation mean free path, $l_\mu$. Such systems bear analogies to the hydrodynamics of dilute fluids, where the interparticle interactions are low. Electron-electron interaction does not deteriorate the bulk mobility, since it conserves the total momentum of the system. However, the hydrodynamic effects induced by *e-e* scattering exert an important influence on transport properties in mesoscopic systems, where the lateral device dimensions are smaller than $l_\mu$. Mesoscopic devices fabricated on high-mobility 2DESs, where $l_\mu$ can easily exceed the device dimensions [5-9], and where $l_{ee}$ can be similarly long, thus often display behavior reminiscent of the hydrodynamics of dilute fluids [1]. In this work we describe a carrier pumping effect observed when a beam of ballistic electrons in a 2DES sweeps past an aperture, reminiscent of the Bernoulli pressure-drop effect in classical fluids. We will also point out qualitative differences between the carrier pumping effect in a degenerate electron system, and the Bernoulli effect.

## 2. Experimental results

We have performed charge transport experiments under externally applied magnetic fields on a 2DES patterned in the geometry depicted in Fig. 1. The 2DES is contained in a GaAs/AlGaAs heterostructure. The darker lines in Fig. 1 denote wet-etched regions, depleted of carriers and functioning as electrostatic barriers for the electrons. The apertures (*a*, *b* and *c*) feature a nominal width of 0.6 µm, whereas the actual conducting width $L$ is smaller due to side depletion, and is estimated at $L \approx 0.4$ µm. In our measurements, a current is drawn between aperture *a* and a faraway contact (not depicted), and either the current or the voltage induced in apertures *b* and *c* are measured as function of a perpendicularly applied magnetic field $B$. The magnetic field is utilized to modify the electron trajectories by bending the latter into cyclotron orbits (circular for electrons in GaAs). As depicted in Fig. 1, over a range of $B$, the semi-classical electron trajectories emitted from *a* will graze the cavity containing apertures *b* and *c*. At these values of $B$, an extremum is observed in the voltage or current recorded in the leads attached to apertures *b* and *c* (Fig. 2). The extremum possesses a sign consistent with carrier extraction from the apertures.

The GaAs/AlGaAs triangular well heterostructure contains a high-mobility 2DES, 700 Å below the surface. The mesoscopic geometry was patterned by electron beam lithography and shallow wet etching, using PMMA as an etching mask. A Cr/Au gate covers the entire sample, allowing us to change the areal density, $N_s$. For the data presented here, $N_s = 4.6 \times 10^{15}$ m$^{-2}$, at a mobility $\boldsymbol{m} = 180$ m$^2$/Vs. The momentum relaxation mean free path, $l_\mu$, then is calculated at 20 µm. Measurements were performed at temperatures $T = 330$ mK and 1.3 K, utilizing mostly low excitation currents (100 nA) to avoid electron heating. Some experiments, however, were conducted both with AC lock-in techniques and in DC mode at various excitations, to ascertain linearity of the response. Four-contact configurations were utilized for all experiments.

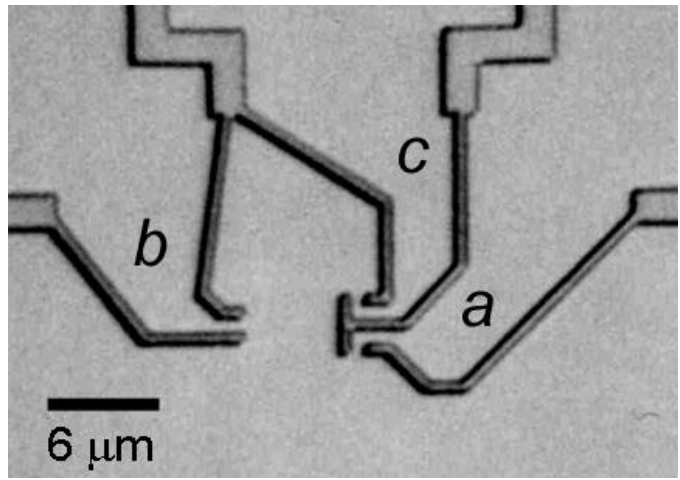

Figure 1: Optical micrograph of a mesoscopic structure on a 2DES, in which the pumping effect is observed. The dark lines denote wet-etched regions from which carriers are excluded. Carriers are injected from aperture *a*, and under a perpendicular magnetic field, sweep past the cavity into which apertures *b* and *c* terminate.

Figure 2 (solid line) shows the potential $V_b$ recorded in the lead attached to aperture *b* (*vs.* a faraway Ohmic contact) when a current $I_a$ of 100 nA is applied between aperture *a* and a fourth and faraway Ohmic contact, as a function of applied *B*. The data was obtained using low-frequency AC lock-in techniques, at $T = 330$ mK. In this AC measurement, a negative value indicates that the potential in contact *a* is of opposite polarity to that in contact *b*, referred to the faraway contacts, whereas a positive sign indicates the same polarity. The negative peak, with minimum at $B = 0.027$ T, then occurs when a surplus of electrons in the injecting aperture *a*, leading to an electron current $I_a$ injected from *a*, induces a lack of electrons in aperture *b*. The negative $V_b$ signal thus corresponds to carrier extraction from *b*. At $B = 0.027$ T, the cyclotron orbit diameter equals 8.3 µm, resulting in an orbit that straddles the cavity into which apertures *b* and *c* terminate. Hence, we expect the entire cavity to adopt a potential consistent with electron extraction. Indeed, the potential $V_c$ measured at aperture *c* reflects this fact, as the dotted line in Fig. 2 shows. The potential at aperture *c* also displays a minimum at $B \approx 0.027$ T. For $B > 0.027$ T the cyclotron orbit enters the cavity and locally increases the electron density, leading to a rapidly rising signal, as evidenced from the data in Fig. 2. For still higher magnetic fields, the cyclotron orbit diameter becomes comparable to the aperture openings, and ceases to determine the ballistic magnetovoltage phenomena.

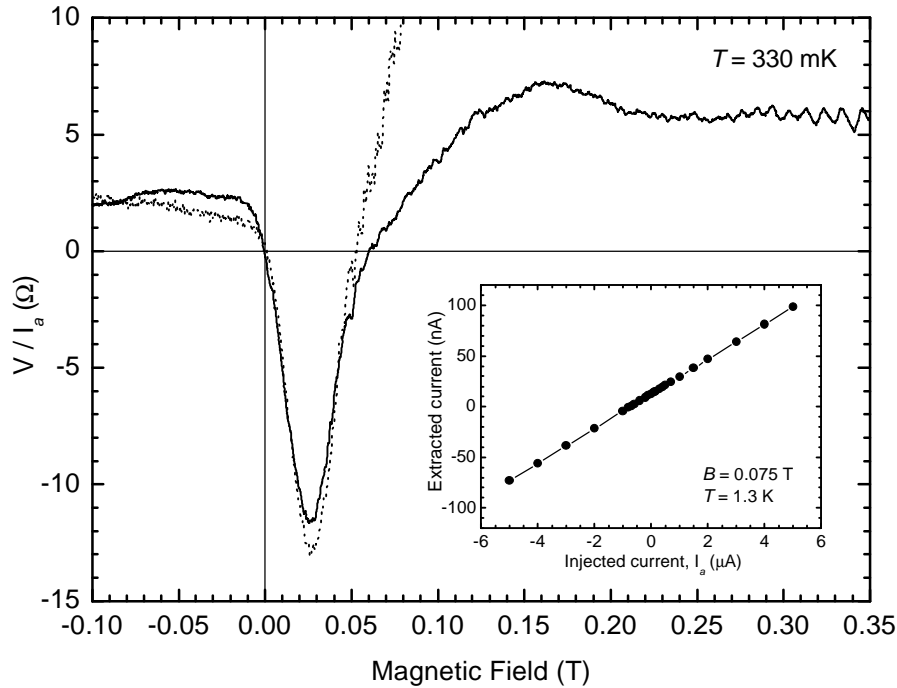

Figure 2: The potential measured at aperture *b* (solid line) and *c* (dotted line), referred to a faraway contact, as function of the applied perpendicular magnetic field and normalized to the current injected from aperture *a* (AC measurement). Inset: linear dependence of the current measured through aperture *b* as a function of the current injected from aperture *a*. A current offset is present due to the voltage bias of aperture *b* (DC measurement).

Further experimental aspects deserve mention. Measurement of the current induced through aperture $b$ have been conducted, both in DC and using AC lock-in techniques, at $T =1.3$ K (inset in Fig. 2). Although complications arise due to the sensitive voltage biasing requirements necessary not to upset the potential landscape under such measurements, the current measurements confirm our conclusion of electron extraction. We moreover performed the voltage and current measurements under varying excitation levels in order to investigate the linearity of the pumping phenomenon, and to investigate the effect of electron heating. No deviations from linearity were observed over current biases of $\pm 5$ µA (inset in Fig. 2).

Some features of the data remain unexplained. Firstly, the smaller feature at $B \approx 0.050$ T persistently appears in the voltage traces for both apertures $b$ and $c$. The origin of this feature (where the cyclotron orbit diameter equals 4.5 µm, ) is so-far unknown. Secondly, the difference in potential appearing between apertures $b$ and $c$ under current injection from aperture $a$, and as $B$ is swept, hitherto has escaped explanation and requires further analysis.

It is interesting to contrast our observations with the previously observed effect of negative ballistic bend resistance [7,10,11]. Assuming electron injection, in the geometry giving rise to the negative bend resistance, an aperture opposite the injector receives a ballistic beam of electrons, and assumes a potential commensurate with the excess charge. In our geometry, no aperture is positioned opposite the injector, and instead, apertures $b$ and $c$ experience a lack of electrons.

## 3. Theoretical description

To model the electron extraction effect due to the ballistic beam, we will consider a geometry shown in the inset of Fig. 3. The electron beam is injected through an aperture in the vicinity of the origin $r = 0$, where $r = (x,y)$ is the in-plane coordinate. We can neglect quantization of the conductance through the apertures, since ~ 20 transverse channels are occupied for an aperture width of 0.4 µm [12,13]. The problem will be solved by using the Boltzmann equation with an $e$-$e$ collision integral. At low temperatures, the distribution function of non-equilibrium electrons is nonzero only in the small vicinity of the Fermi level and is written as $f(e,q) = -df_0/de\, F(q)$, where $f_0$ is the equilibrium Fermi function, and $e$ and $q$ are the electron energy and the angle between the electron velocity and the direction $+x$. The boundary condition at the line $x = 0$ and outside of the aperture describes the elastic collision of an electron with an ideal border, $F(y;q) = F(y;p\text{-}q)$. On the line $x = 0$ and at small coordinates $y$, the boundary condition should include the injection effect. Thus, everywhere on the line $x = 0$, the boundary condition reads

$$F(y;\boldsymbol{q}) - F(y;\boldsymbol{p} - \boldsymbol{q}) = g(y;\boldsymbol{q}) - g(y;\boldsymbol{p} - \boldsymbol{q}), \tag{1}$$

where the function $g(y;\boldsymbol{q})$ describes the injection and is nonzero in the interval $-\boldsymbol{p}/2 < \boldsymbol{q} < \boldsymbol{p}/2$.

To describe the current injection, we now introduce symmetric current sources in the linearized Boltzmann equation:

$$\mathbf{v}\frac{\partial F}{\partial \mathbf{r}} - e\mathbf{v}\mathbf{E} = -\frac{F - \bar{F} - 2\cos(\boldsymbol{q})J_x - 2\sin(\boldsymbol{q})J_y}{t_{ee}} + G(y;\boldsymbol{q})\boldsymbol{d}(x) + G(y;\boldsymbol{p}-\boldsymbol{q})\boldsymbol{d}(x),$$
(2)

where $\mathbf{v}$ is the electron velocity, $t_{ee}$ is the $e$-$e$ collision time,

$$\bar{F} = \int_0^{2p} F(\boldsymbol{q})d\boldsymbol{q} / 2\boldsymbol{p},$$

$$J_x = \int_0^{2\pi} F(\varphi) \cos(\varphi) d\varphi / 2\pi,$$

$$J_y = \int_0^{2\pi} F(\varphi) \sin(\varphi) d\varphi / 2\pi,$$

and **E** is the in-plane electric field. Equation 2 was written within the relaxation time approximation. It is easy to show that the solution of Eq. 2 satisfies the boundary conditions (1) if we choose $G(y;\varphi) = v_F \cos(\theta) g(y;\varphi)$, where $v_F$ is the Fermi velocity. For **E**, we will use a local, flat capacitor approximation, which is valid in the structure covered by the metallic top gate, just like in our experiments. So, $\mathbf{E} = -4\pi e\, d / \varepsilon_{sem} (d\delta n/dr)$, where $\delta n(r)$ is the non-equilibrium electron density, $d$ is the distance between the 2DES and the metal gate, and $\varepsilon_{sem}$ is the dielectric constant of the semiconductor.

Equation 2 was solved analytically by Fourier transformation in the whole 2D plane [14]. After the reverse transformation, we consider the function $F(x;y;\varphi)$ only in the right-hand half of the 2D plane. As injection function in the linear regime, we used

$$g(y;\varphi) = g_0 e^{-y^2/L^2} \delta(\varphi) \delta(\varepsilon - E_F),$$

where $E_F$ is the Fermi energy.

Figure 3 contains the carrier density distribution in the vicinity of the injecting aperture, calculated as outlined above. The non-equilibrium electron density is enhanced inside the ballistic beam, and reduced in the regions flanking the beam (Fig. 3). In Fig. 4, we schematically show the streamlines. In the regions nearby the main electron beam, the currents flow *toward* the beam of injected electrons. Qualitatively, we can understand such a behavior in the following way. The injected electrons moving from the aperture scatter the background Fermi-sea carriers towards the right. Numerical calculations show that this happens not in the main beam, where the density of excess electrons is high, but rather on the sides of the electron beam. The current flowing in the *y*-direction tends to compensate the lack of electrons, and leads to carrier extraction from the regions surrounding the beam.

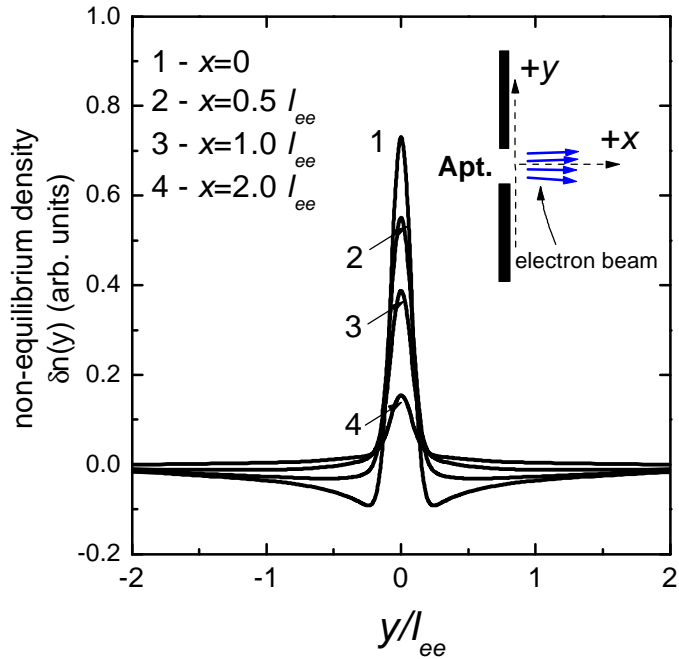

Figure 3: Calculated non-equilibrium density as a function of the $y$-coordinate at several $x$-positions. In the calculation, the aperture width $L = 0.1\, l_{ee}$. Insert: the geometry of injection.

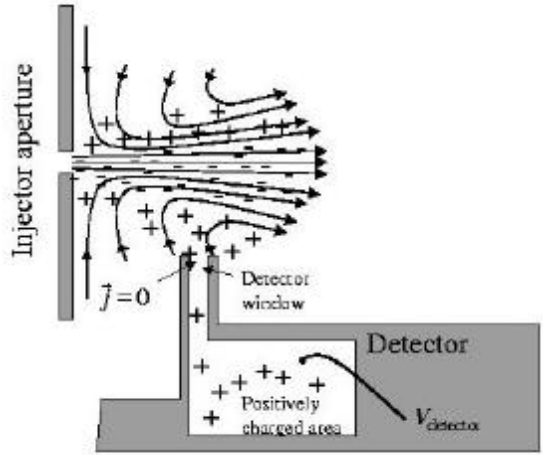

Figure 4: The streamlines and density distribution (schematically). The detector with window and reservoir is introduced as a probe for the potential near the main beam of electrons.

The density calculations in Fig. 3 also explain the experimentally observed hydrodynamic pumping voltage (Fig. 2), as follows. We assume that the aperture of a detector (*b* or *c*, or the cavity, in Fig. 1) is placed in the vicinity of the electron beam (Fig. 4). The potential in the presence of a non-equilibrium density δ*n* can be expressed as $V \approx (4\pi ed / e_{sem}) \delta n$, reflecting the fact that a local net charge density will lead to a local potential of the same sign. The electron density is depressed at the detector window if the latter is in the region where δ*n* < 0, resulting in a positive charge density and positive detector potential. This is what we observe experimentally. Since during electron injection, the injector aperture (*a* in Fig. 1) has the opposite, negative, potential, the sign of the measured lock-in signal will be negative, as indeed observed in Fig. 2. The linearity of the phenomenon ensures that the argument holds identically in the positive AC lock-in cycle. Linearity is experimentally observed (inset in Fig. 2). Theoretically, linearity follows from the fact that $V \sim \delta n$, and $\delta n \sim I_a$, the ballistic beam current from the injector aperture. In voltage measurement mode, the detector can be considered as a closed reservoir because in steady state no current flows from it, and this fact can be utilized to further clarify the generation of a potential. Suppose we turn the injected current $I_a$ on at time *t* = 0. In the region of the detector window, the pumping effect extracts electrons from the detector. After some time the system reaches a steady state, and the net current through the detector window vanishes. Hence a countercurrent of electrons into the detector must be generated, by a positive potential on the detector lead. Our theoretical scheme can also be used to pump current through the detector window. To achieve this, we can connect the detector reservoir with the Fermi sea in the right-hand side of the 2D plane, including a resistor in the external circuit. Our picture will hold if the external resistor is sufficiently large and the pumped current can be considered as a weak perturbation. In our experiments, we realized both the scheme in which the detector lead does not carry a current (voltage measurement) and the scheme in which the detector current is allowed to flow (current measurement).

It is interesting to compare our results with hydrodynamic effects in classical liquids. Of course, the local non-linear equations of fluid mechanics are qualitatively different from the non-local kinetic Boltzmann equation. For example, the Bernoulli pumping effect, resulting from the spatial variation of local speed in the moving liquid, is *quadratic* in terms of fluid speed and *non-linear*. Hence the Bernoulli pumping effect is not changed if we reverse the current. In our case, the pumping force is linear in terms of applied voltage or current and changes its sign as we reverse the current flow. This linearity is a consequence of the Fermi statistics of electron. At low temperature, we can consider the flow equally in terms of electrons or of holes near the Fermi surface.

## 4. Conclusions

To summarize, we have described a hydrodynamic pump in a Fermi liquid. We show that the Coulomb interaction in an injected ballistic beam leads to attractive forces and to a carrier pumping effect in a degenerate electron system.

## 5. Acknowledgements

We thank M. Shayegan (Princeton University) for the high-mobility GaAs/AlGaAs heterostructures. J. J. H. acknowledges support from the National Science Foundation, under grant DMR-0094055. A. O. G. acknowledges support from the Condensed Matter and Surface Science Program at Ohio University.